%% file: paper.tex
\documentclass[iop]{emulateapj}
\usepackage[T1]{fontenc}
\usepackage{microtype}
\usepackage{amsmath}
\usepackage{listings}
\usepackage[colorlinks,urlcolor=blue,citecolor=blue,linkcolor=blue]{hyperref}

\slugcomment{Submitted to PASP, 2012 Mar 1}
\shorttitle{Rapid Development with MIRIAD \& Python}
\shortauthors{Williams {\it et al.}}

\newcommand\citeeg[1]{\citep[e.g.,][]{#1}}
\newcommand\apx{\ensuremath{\sim}}
\newcommand\fnu[1]{\footnote{\url{#1}}}
\newcommand\mirpy{\textsf{miriad-python}}
\newcommand\git{\textsf{git}}
\newcommand\miriad{\textsf{miriad}}
\newcommand\mirtask{\textsf{mirtask}}
\newcommand\mirexec{\textsf{mirexec}}
\newcommand\uv{$u$\mbox{-}$v$}
\newcommand\cursha{\textsf{aff0a3cb\-4d47d030\-426cf9e3\-6475e4bc\-9ae1816f}}
\newcommand\currelvers{0.6}
\newcommand\curreldate{2011~April~28}
\newcommand\currelsha{\textsf{acbfa731\-c5da07c0\-270b03e9\-e93c11f4\-c9b7a4d4}}

\input figaliases.tex

\begin{document}

\lstset{
  language=Python,
  basicstyle=\small\sffamily,
  showstringspaces=false,
  columns=fullflexible,
  xleftmargin=12pt,
}

\title{Rapid Development of Interferometric Software Using MIRIAD and
  Python}

\author{Peter~K.~G.~Williams, Casey~J.~Law, Geoffrey~C.~Bower}
\affil{Department of Astronomy, B-20 Hearst Field Annex~\#~3411,
  University of California, Berkeley, CA 94720-3411, USA}
\email{pwilliams@astro.berkeley.edu}

\begin{abstract}
  New and upgraded radio interferometers produce data at massive rates
  and will require significant improvements in analysis techniques to
  reach their promised levels of performance in a routine
  manner. Until these techniques are fully developed, productivity and
  accessibility in scientific programming environments will be key
  bottlenecks in the pipeline leading from data-taking to research
  results. We present an open-source software package, \mirpy, that
  allows access to the MIRIAD interferometric reduction system in the
  Python programming language. The modular design of MIRIAD and the
  high productivity and accessibility of Python provide an excellent
  foundation for rapid development of interferometric
  software. Several other projects with similar goals exist and we
  describe them and compare \mirpy\ to them in detail. Along with an
  overview of the package design, we present sample code and
  applications, including the detection of millisecond astrophysical
  transients, determination and application of nonstandard calibration
  parameters, interactive data visualization, and a reduction pipeline
  using a directed acyclic graph dependency model analogous to that of
  the traditional Unix tool \textsf{make}. The key aspects of the
  \mirpy\ software project are documented. We find that
  \mirpy\ provides an extremely effective environment for prototyping
  new interferometric software, though certain existing packages
  provide far more infrastructure for some applications. While
  equivalent software written in compiled languages can be much faster
  than Python, there are many situations in which execution time is
  profitably exchanged for speed of development, code readability,
  accessibility to nonexpert programmers, quick interlinking with
  foreign software packages, and other virtues of the Python language.
\end{abstract}


\section{Introduction}
\label{s:intro}

Advances in the fields of digital computing and commercial wireless
communication have fueled an explosion of innovation in radio
interferometry. New and upgraded facilities such as the Allen
Telescope Array \citep[ATA;][]{theata}, the Low-Frequency Array
\citep{thelofar}, the Precision Array for Probing the Epoch of
Reionization \citep[PAPER;][]{thepaper}, the Murchison Wide-Field
Array \citep[MWA;][]{themwa}, the Karl G. Jansky Very Large Array
\citep[formerly EVLA;][]{thekarl}, the Westerbork Synthesis Radio
Telescope (WSRT) Apertif project \citep{theapertif}, the Australian
Square Kilometer Array Pathfinder \citep{theaskap}, and MeerKAT
\citep{themeerkat} have sophisticated designs including
large-number-of-small-dishes architectures, multipixel or phased-array
feeds, wide bandwidths, and phased array substations. Several of these
facilities are pathfinders for the proposed Square Kilometer Array
\citep[SKA;][]{theska}. They all aim to push the limits of
interferometric techniques to image large fields of view with
unprecedented spatial, spectral, temporal, and polarimetric fidelity,
at unprecedented data rates.

%{\bf [This is placeholder text to work around a bug in hyperref
%    triggered by the next paragraph. It aborts if you use two-column
%    mode and have a citation that spans multiple pages.]}

Achieving these goals will require substantial amounts of new software
to process the data coming out of these facilities. The data rates
present a challenge in and of themselves: SKA-class facilities will
require exaflop-scale computing \citep{ch10b}. Another challenge is
the development of the necessary new algorithms, which is certain to
require extensive experimentation. Techniques already under
investigation include $w$-projection \citep{cgb08}, $A$-projection
\citep{bcgu08}, multi-scale multi-frequency CLEAN \citep{rc11},
delay/delay-rate filtering \citep{pb09}, space-alternating generalized
expectation (SAGE) maximizing calibration \citep{kyz+11}, generalized
measurement-equation-based instrumental modeling \citep{ns10},
scale-invariant rank detection of radiofrequency interference
\citep[RFI;][]{ovdgr12arxiv}, subspace-tracking RFI mitigation
\citep{eh02}, visibility stacking \citep{hgm11}, bispectral pulse
detection \citep{lb11arxiv} and improved sourcefinding tools
\citep{w12arxiv}, to name a few. Until the next generation of
algorithms is thoroughly explored, the quality of many results will be
set not by the capabilities of observatory hardware but by the
sophistication of the reduction pipeline that can be brought to bear
before publication: {\it software-limited} science.

Although improvements in software development efficiency are always
desirable, they're particular salient now as the next generation of
radio interferometers comes online. We discuss the efficiency of a
programming environment in terms of {\it productivity}, which we take
to measure the amount of useful functionality that a programmer can
implement per unit time, and {\it accessibility}, which measures how
much effort it takes for non-experts to begin successfully working
within the environment without hand-holding. Although many studies
attempt to quantify these metrics \citep[e.g.,][and references
  therein]{p11}, our discussion will remain qualitative. These
attributes are relevant in situations in which developer resources are
constrained: a more productive environment allows individual
developers to accomplish more, while a more accessible environment has
a lower barrier to entry from informal contributors.

Interpreted, dynamic programming languages can provide such an
environment. While several of these exist, the language
Python\fnu{http://python.org/} in particular has seen a broad uptake
in the astronomical community over the past decade
\citeeg{thepyfits,thepyraf,bam+03,klrc06,mbi07}. It is intended to be
easy to learn and offers conveniences including object-oriented
programming, lambda expressions, exception handling, and an enormous
software ecosystem, including interactive interpreters \citep[e.g.,
  IPython;][]{theipython}, visualization tools \citep[e.g.,
  \textsf{matplotlib};][]{thematplotlib}, file format interfaces
\citep[e.g., \textsf{pyFITS};][]{thepyfits}, database interfaces
(e.g., SQLite\fnu{http://sqlite.org/}), statistics routines, web
service support, and so on. The existence of this ecosystem and the
rapid rise in the popularity of Python in the astronomical community
in particular speak to its productivity and accessibility.

The programming environments of ``classic'' astronomical software
packages tend to be intimidating and frustrating in comparison. Two of
the major traditional radio interferometric reduction packages, MIRIAD
\citep[Multichannel Image Reconstruction, Analysis, and
  Display;][]{themiriad} and AIPS \citep[Astronomical Image Processing
  System;][]{theaips}, are largely implemented in FORTRAN-77, and in
both cases the process to go from source code to usable installed
executable code is complicated and fragile. In light of the clear need
for improved interferometric algorithms, it should be no surprise that
a variety of projects have sought to build on or replace these systems
with more modern ones. Perhaps the most prominent such undertaking is
the Common Astronomy Software Applications \citep[CASA;][]{thecasa},
the successor to AIPS. Other packages include Obit \citep{theobit},
AIPY (Astronomical Interferometry in
Python\fnu{http://purl.org/net/pkgwpub/aipy}), MeqTrees \citep{ns10},
and several more narrowly-targeted binding layers \citep[e.g.,
  \textsf{pyramid};][]{thepyramid}.

To this list, we add \mirpy, a creatively-named package exposing
MIRIAD tasks and subroutines in Python. Its design and philosophy are
discussed (\S\ref{s:design}) and compared those of related projects
(\S\ref{s:compare}), including the ones mentioned above. We then
describe the implementation of \mirpy\ (\S\ref{s:impl}) and provide
some very brief examples of its use (\S\ref{s:examples}). Some
applications in which it has been used are presented, including the
detection of millisecond astrophysical transients, determination and
application of nonstandard calibration parameters, interactive data
visualization, and pipeline processing (\S\ref{s:apps}). We document
the nature of \mirpy\ as a software project (\S\ref{s:project}),
discuss some of its performance characteristics (\S\ref{s:perf}), and
finally summarize (\S\ref{s:summary}).

\section{Design Considerations in \mirpy}
\label{s:design}

The \mirpy\ project is intended to allow convenient access to MIRIAD
tasks, datasets, and subroutines. Although it aims to ease the rapid
development of new interferometric algorithms, it does not provide any
nontrivial algorithms itself. As such, the focus of \mirpy\ is very
narrow: it provides the best possible interfaces for access to MIRIAD
infrastructure and is as flexible as possible regarding what is done
with it. To use terminology dating back to at least the design of the
X~Windows system, it provides {\it mechanism} but not {\it policy}
\citep{thexwindows}. We believe that the wide range of applications
described in \S\ref{s:apps} is evidence of the power of this approach.

Although work on \mirpy\ was initially inspired by a practical desire
for more efficient scripting of MIRIAD reductions, the MIRIAD package
is a good fit to the larger goals of the \mirpy\ project. MIRIAD
offers its wide variety of tools and algorithms in a modular task
architecture, and its data formats are simple and efficient. In
particular, it is fairly straightforward to (ab)use the streaming
MIRIAD \uv\ data format for novel applications (see, e.g., AIPY,
\S\ref{s:highlevel}). Other packages are built upon MIRIAD for similar
reasons \citeeg{thedrpacs,themis}. MIRIAD is routinely used to process
data from facilities including the ATA, the Berkeley-Illinois-Maryland
Association array \citep[BIMA;][]{thebima}, the Combined Array for
Research in Millimeter/submillimeter Astronomy
\citep[CARMA;][]{thecarma}, the Australia Telescope Compact Array
(ATCA), and the Submillimeter Array \citep[SMA;][]{thesma}. MIRIAD is
not a monolithic project: there are at least two nontrivially
divergent codebases maintained for use with the ATCA and CARMA, with
some sharing of modifications between the them. \mirpy\ is referenced
to the CARMA MIRIAD codebase. Recent work on MIRIAD includes complete
support for 64-bit file offsets and pointers, allowing imaging of
arbitrarily large datasets; integration of the \textsf{wcslib} library
\citep{thewcslib} for more comprehensive support of coordinate
manipulations; and of course bug fixes, new features, and
documentation improvements.

When creating a package such as \mirpy\ that interfaces with a
lower-level one, one must decide how closely to hew to the APIs
(application programming interfaces) of the original package. In
\mirpy, the APIs have been significantly reworked: they are
object-oriented and aim to take full advantage of builtin language
features. When they meet the goal of providing mechanism rather than
policy, substantial new features are implemented in the Python
layer. One example is approximate cryptographic hashing of
\uv\ datasets, which can be used to quickly and fairly robustly check
for modifications to the data regardless of file modification
times. (This is useful in dependency-tracking pipelines,
cf. \S\ref{s:atapipeline}.) To write useful programs, \mirpy\ authors
must be familiar with the semantics of MIRIAD datasets or tasks, and
good knowledge of Python is helpful, but they do not need to
understand the design of the MIRIAD subroutine library.

Another consideration for authors is whether to link to a separate
installation of the lower-level software or to compile and install
their own version of it, bundling a copy of its source code with their
own. We have chosen to go the former route with \mirpy. Although
\mirpy\ offers very different APIs and new features, we concieve of it
as fundamentally a layer above MIRIAD and not a standalone package: if
the user has customized their MIRIAD installation in some way, for
instance, it's appropriate for \mirpy\ to reflect that customization,
and not override it with its own copy of the subroutine
library. Another reason is that MIRIAD is still evolving, and tracking
changes from an upstream codebase to a forked copy can be tedious and
error-prone. Another is that \mirpy\ is a general-purpose interface to
not just the MIRIAD subroutine library but its task collection, so the
entire, substantial, MIRIAD codebase would need to be
bundled. Finally, we expect that \mirpy\ users will be existing MIRIAD
users who are likely to already have installed MIRIAD on their
systems, so the convenience of bundled source code is lessened.

To touch on a broader philosophical issue, it may be argued that
accessibility is not a desirable feature when it comes to
interferometric software. Interferometric analysis tends to be subtle,
and most people write buggy code (see \S\ref{s:qa}). On the other
hand, scientific research is a fundamentally creative and exploratory
process often requiring new or improved techniques, and that is
clearly the case in the domain of interferometry at present. The
tension here is a variation on the well-explored ``cathedral versus
bazaar'' theme first described by \citet{r99}. Considering the amount
of algorithmic development necessary to fully exploit next-generation
radio interferometers, we take the stance that the empowerment of
interferometric software users is a good thing.

\section{Comparisons to Related Packages}
\label{s:compare}

As alluded to in \S\ref{s:intro}, there are several other efforts
underway to provide developer-friendly environments for producing
interferometric software. In this section we attempt to situate
\mirpy\ among them. We consider ``bindings'', which do not provide
substantial new algorithms, and ``high-level'' packages, which do.

\subsection{Bindings}

Several other packages expose MIRIAD functionality in the Python
language, but we found them unsatisfactory. The \textsf{pyramid}
package \citep{thepyramid}, providing a module named \textsf{Miriad},
includes similar functionality to that of \textsf{mirexec} in a
somewhat less structured manner. The modules \textsf{mirlib} (via the
WSRT) and \textsf{miriadwrap} both bind the MIRIAD subroutine
library. Unlike \mirpy, they provide a fairly direct mapping to the
MIRIAD subroutines and do not extensively ``Pythonify'' the API. These
packages are less mature than \mirpy\ and do not appear to be under
active development. None of them includes substantial documentation,
high-level convenience features, or an integrated system for invoking
both MIRIAD tasks and subroutines.

There also exist several other packages that appear to share the same
motivations as \mirpy\ but involve different technologies. Other
dynamic languages can be used. For instance, MIRIAD functionality can
be accessed in Ruby\fnu{http://www.ruby-lang.org/} with the
\textsf{MIRIAD-Ruby}\fnu{http://purl.org/net/pkgwpub/miriad-ruby} or
\textsf{mirdl}\fnu{http://purl.org/net/pkgwpub/mirdl} packages. The
two packages, written by the same author, cover the link-versus-bundle
tradeoff discussed above: \textsf{MIRIAD-Ruby} includes a portion of
the MIRIAD source code, while \textsf{mirdl} links to a separate
installation of the full MIRIAD libraries. Although it is beyond the
scope of this paper to compare the merits of the languages, we note
that Ruby has seen less uptake in the astronomical community than
Python but is starting to make inroads \citeeg{lbtv02,fmg07,gkkm+10}.

Other packages build on top of different interferometry
frameworks. ParselTongue \citep{klrc06} provides a Python interface to
AIPS, building on top of parts of the Obit package \citep{theobit},
described in the next subsection. The \textsf{pyrap}
project\fnu{http://purl.org/net/pkgwpub/pyrap} provides a Python
interface to the core C++ support libraries of CASA.

In many cases, the choice of the appropriate binding will be
constrained by external limitations in either the high-level language
or the underlying interferometry package. Our reasons for preferring
Python and MIRIAD, respectively, are sketched in \S\ref{s:intro} and
\S\ref{s:design}. We emphasize the simplicity and efficiency of
MIRIAD's data formats as being important factors in facilitating
algorithmic experimentation. Another distinction between \mirpy\ and
alternative bindings is its rate of development. Even in otherwise
stable and mature packages, documentation improvements and subtle
bugfixes tend to require fairly frequent modifications. In the
calendar year of 2011, there were 5 commits to the \textsf{mirdl}
version control system (VCS), 10 commits to the \textsf{pyrap} VCS,
and 110 commits to that of \mirpy. The last known update to
ParselTongue was in 2010 and the last known update to \textsf{pyramid}
was in 2008. Finally, we believe that \mirpy\ acquits itself well when
it comes to the subjective factors of API cleanliness, code
readability, and code quality.

\subsection{High-Level Packages}
\label{s:highlevel}

The projects described in this subsection are more ambitious than
\mirpy\ and include significant functionality along with Python-based
environments: they provide policy, as well as mechanism
(cf. \S\ref{s:design}). While the packages described in the previous
subsection are intended to provide suitable bases for development for
a wide range of applications, the ones described below may not be
appropriate for certain applications, depending on both technical and
architectural factors. All of these packages are actively developed.

CASA \citep{thecasa} is a substantial package including high-level
Python interfaces to lower-level routines implemented in C++ and
FORTRAN. The first iteration of CASA, known as AIPS++
\citep{theaipspp}, included flexibility, ease of programming, and
modifiability in its mission statement \citep{aipsppsc93}. We find
that these are difficult to achieve in practice because most CASA
functionality is implemented in fairly large C++ modules, with the
built-in Python interfaces providing mostly coarse-grained access to
data and algorithms; the combination of the core CASA libraries with
\textsf{pyrap} is more friendly to Python development. One minor
inconvenience is that CASA bundles its own Python interpreter, so that
site-specific packages accessible to the system interpreter sometimes
need to be rebuilt and reinstalled in order to be accessible to CASA's
version. CASA's imager is quite flexible and can combine multiple
advanced algorithms, for instance simultaneous multi-scale,
multi-frequency deconvolution \citep{rc11}.

Obit is billed as ``a development environment for astronomical
algorithms'' \citep{theobit}. Although many of its fundamental
features are general-purpose, Obit's main emphasis is radio astronomy,
and in particular it includes several algorithms important to
low-frequency radio interferometry. The package includes a set of base
libraries implemented in C, a set of tasks built on top of them, and a
Python layer for interacting with both of these. This layer is based
on ParselTongue, although the two projects remain separate
(W.~D.~Cotton, priv. comm.). Two aspects of the Obit design are of
particular note. Firstly, Obit uses AIPS and FITS file formats for
data storage, so it interoperates seamlessly with certain existing
packages and developer effort need not be spent reimplementing
existing tools. Secondly, the Obit libraries are architected to allow
multithreaded execution, and this capability has been explored during
Obit's development. In one example problem, the time taken to generate
a complex image was reduced by a factor of 6.5 when the number of CPU
cores used for computations was increased from one to twelve
\citep{cp10}.

The AIPY system represents a sharper break from tradition than those
previously mentioned. It is a standalone package implementing many
advanced synthesis and calibration techniques in Python to meet the
needs of the low-frequency dipole array PAPER. Its design emphasizes
flexibility and is Python-centric: the primary interfaces are the
Python ones, and code from preexisting packages such as HEALPix
\citep{thehealpix} and XEphem \citep{thexephem} is liberally borrowed
to provide fast, well-tested implementations of many features. AIPY
can use the MIRIAD format for \uv\ data storage, and hence includes an
internal Python binding of the relevant portions of the MIRIAD
subroutine library. A particularly interesting strength of AIPY is its
infrastructure for creating maps of the whole sky at once.

The MeqTrees project \citep{ns10} also uses Python to provide a
productive environment for exploring new interferometric algorithms
using the measurement equation formalism \citep{hbs96,s11}. Its
input/output (I/O) and astronomical infrastructure are built on the
core CASA libraries and \textsf{pyrap}, with certain numerical
routines implemented in C++. MeqTrees provides a substantial
infrastructure for developing and applying sophisticated calibration
methodologies with numerous visualization features. As an example of
its power, MeqTrees has been used to generate a noise-limited radio
astronomical image of the source 3C\,147 with a dynamic range of
$1.6\times10^6$ \citep{ns10}.

\section{Implementation}
\label{s:impl}

\begin{figure}[htbp]
\plotone{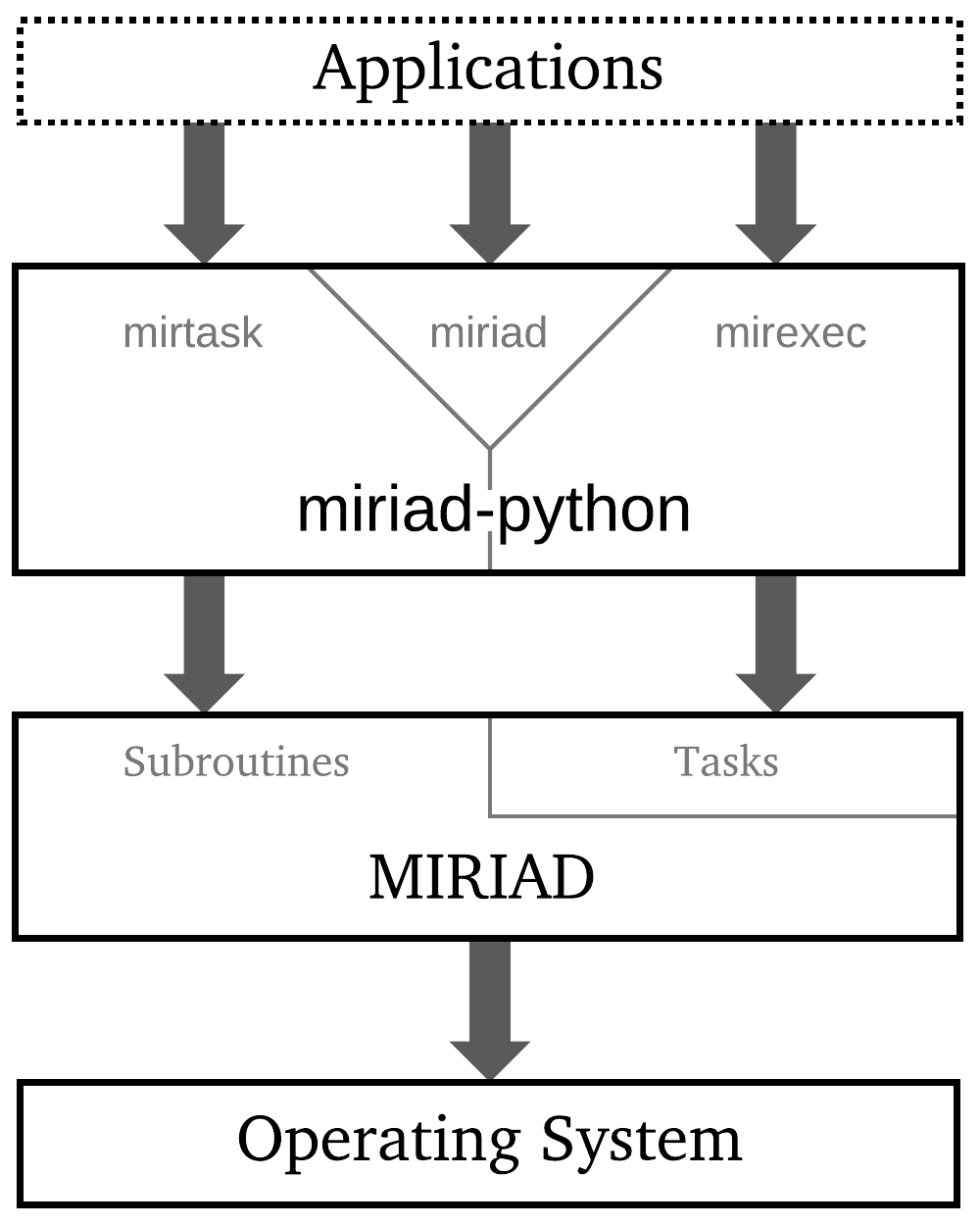}
\caption{Diagram of the relationships between different components of
  the \mirpy\ system. Applications use the three \mirpy\ modules,
  which in turn access the tasks and subroutines in MIRIAD, which
  interact with the base operating system. Both applications and
  \mirpy\ use the NumPy module (not depicted) for numerical array
  operations.}
\label{f:blockdiagram}
\end{figure}

\mirpy\ provides three Python modules: \mirtask, \mirexec, and
\miriad. The first of these uses C and FORTRAN-77 based extension
modules to interface with the MIRIAD subroutine library, providing
most of the low-level functionality that makes
\mirpy\ useful. Numerical functionality is provided by
NumPy\fnu{http://numpy.scipy.org/} and linking with FORTRAN-77 is
accomplished using \textsf{f2py} \citep{thef2py}. The second module,
\mirexec, is Python-only and provides a uniform framework for
executing MIRIAD tasks inside the Python language. The third, \miriad,
is Python-only and provides a simple abstraction for referencing and
manipulating MIRIAD datasets, with hooks into \mirtask\ and
\mirexec\ to provide convenient access to certain common
operations. The three modules are loosely-coupled such that they do
not actually depend on each other, although most nontrivial tasks end
up using functionality from all three modules. The relationships of
the APIs are diagrammed in Figure~\ref{f:blockdiagram}.

As mentioned above, the MIRIAD APIs are significantly reworked and
expanded. In most cases, Python's runtime type information is used to
automatically choose data types when performing lowlevel I/O. Because
these types are important for task interoperability, they must be
specified explicitly when writing new data. MIRIAD's internal errors
are mapped into Python exceptions using a mechanism based on the C
library functions \textsf{setjmp} and \textsf{longjmp}.

\mirpy\ also provides a program that can serve as a help viewer for
either existing MIRIAD tasks or programs written in Python, in the
latter case using standard Python ``documentation strings'' written in
the standard MIRIAD documentation format. Tasks written in Python can
thus seamlessly merge with traditional compiled MIRIAD tasks from the
user standpoint.

\section{Example Code}
\label{s:examples}

Although it is impractical to provide extensive code examples in this
paper, in this section we give a few brief ones. These are necessarily
simplistic and do not demonstrate the most interesting capabilities of
\mirpy. As discussed below (\S\ref{s:docs}), the \mirpy\ source
distribution contains longer samples\footnote{Currently browseable
  online at
  \url{https://github.com/pkgw/miriad-python/tree/master/examples}.}.
We assume a familiarity with typical MIRIAD usage and the Python
language. Many valuable coding practices are ignored below for the
sake of concision.

MIRIAD tasks can be executed with the \mirexec\ module. The following
example demonstrates the parallel execution of multiple tasks: the
\textsf{launch} method starts a task process, returning a handle to it
(a subclass of Python's \textsf{subprocess.Popen}), and the handle's
\textsf{wait} method pauses the caller until the process
completes. (For I/O-intensive operations as shown here, it may
actually be faster to execute the tasks serially to avoid disk
thrashing.)
\begin{lstlisting}
import sys
from mirexec import TaskUVAver

prochandles = [ ]

for p in sys.argv[1:]:
   taskdesc = TaskUVAver (vis=p, out=p+".avg",
                           interval=10)
   prochandle = taskdesc.launch ()
   prochandles.append (prochandle)

for prochandle in prochandles:
   prochandle.wait ()
\end{lstlisting}
\newpage
The above example is intentionally verbose. To stimulate the reader's
imagination, we provide the following variation that uses Python's
``list comprehension'' syntax and avoids the use of several local
variables:
\begin{lstlisting}
import sys
from mirexec import TaskUVAver

for ph in [TaskUVAver (vis=p, out=p+".avg", interval=10)
          .launch () for p in sys.argv[1:]]:
  ph.wait ()
\end{lstlisting}

The following example demonstrates how the ``header variables'' of
MIRIAD datasets may be accessed using \mirpy. The variable
\textsf{vispath} is some string giving the filesystem path of a MIRIAD
\uv\ dataset. As mentioned in \S\ref{s:impl}, storage types of
variables are detected automatically on read, but must be specified
explicitly on write to enforce consistency with preexisting MIRIAD
tasks.
\begin{lstlisting}
import miriad, numpy as np

vispath = ...
ref = miriad.VisData (vispath)
handle = ref.open ("rw")
ncorr = handle.getScalarItem ("ncorr")
handle.setScalarItem ("mine", np.double, np.pi * ncorr)
\end{lstlisting}

The following example replaces the contents of an image with the
magnitudes of its fast Fourier transform (FFT). This can be used, for
instance, to recover the weights used in the imaging process from the
dirty beam image. The variable \textsf{impath} is analogous to
\textsf{vispath}. The value of the variable \textsf{whichplane}
indicates that the first image plane should be selected.
\begin{lstlisting}
import miriad
from numpy import abs, float32
from numpy.fft import fft2, fftshift, ifftshift

impath = ...
whichplane = [ ]
handle = miriad.ImData (impath).open ("rw")
d = handle.readPlane (whichplane).squeeze ()
d = abs (ifftshift (fft2 (fftshift (d))))
handle.writePlane (d.astype (float32), whichplane)
handle.close ()
\end{lstlisting}

Finally, this example inserts information about the synthesized beam
shape of an image into a preexisting relational database.
\begin{lstlisting}
import miriad, sqlite3

impath = ...
handle = miriad.ImData (impath).open ("rw")
dbpath = ...
db = sqlite3.connect (dbpath)

info = [impath]
for item in "bmaj bmin bpa".split ():
  info.append (handle.getScalarItem (item))

db.execute ("INSERT INTO data VALUES (?,?,?,?)", info)
db.commit ()
db.close ()
\end{lstlisting}
Such functionality would be difficult to achieve using FORTRAN.  We
emphasize that in the interests of simplicity and clarity, we have
avoided some of the most important features of the Python language
such as object orientation, first-class functions, and exception
handling.

\section{Applications}
\label{s:apps}

The fast Python development cycle makes it relatively easy to produce
new algorithms and tools for astronomical data analysis. In this
section we present some example applications of \mirpy.

\subsection{Algorithms for Millisecond Transients}
\label{s:mstransients}

We have investigated with new algorithms to study millisecond
transients in the visibility domain using data from the ATA and VLA
\citep{ljb+11,lb11arxiv}. Since interferometers have not traditionally
operated at this time scale, many interferometric packages lack
features needed to work in this regime, e.g., the ability to
dedisperse visibilities and high-resolution timestamps. We used
\mirpy\ to develop new ways to visualize and search these unusual data
streams for transient radio sources.

\begin{figure}[htbp]
\epsscale{0.8}
\plotone{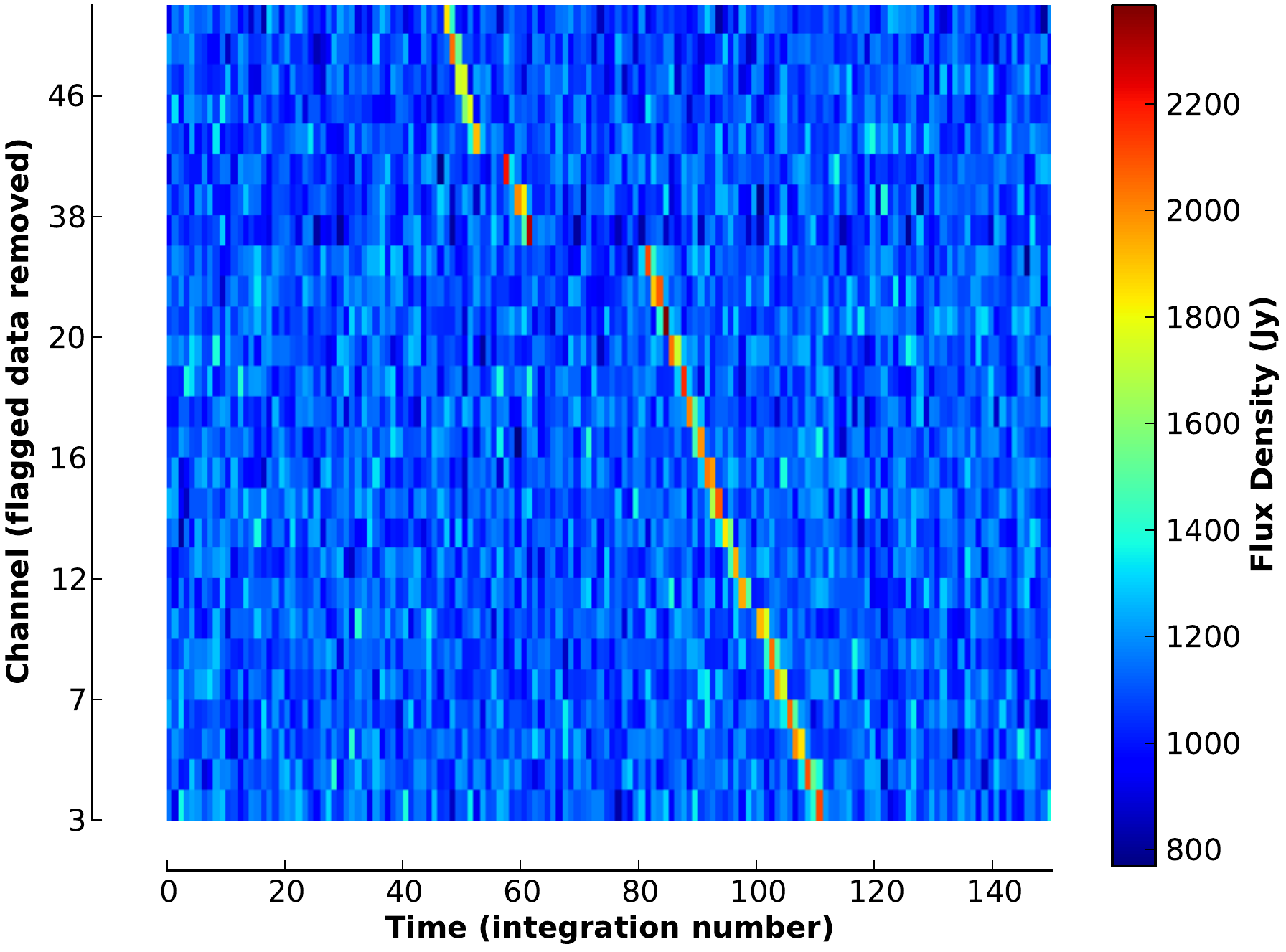}
\plotone{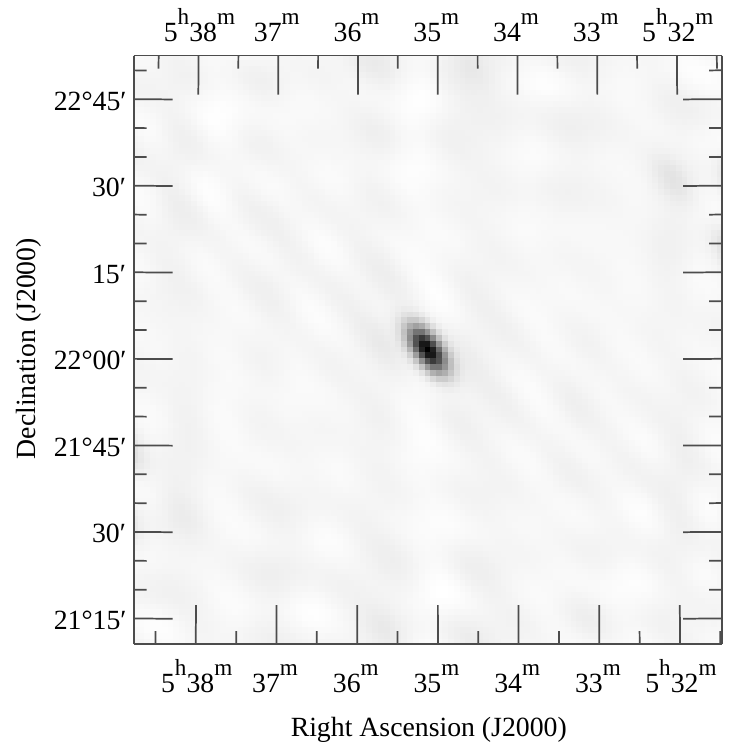}
\epsscale{1}
\caption{{\it Upper panel:} Spectrogram of a pulse from the Crab
  pulsar, summing visibilities coherently using the known location of
  the emitter. Dispersion due to the ISM causes pulse arrival times to
  vary with frequency (channel number). Apparent discontinuities in
  the pulse are due to flagged channels not being rendered (note the
  gaps in the left axis values). {\it Lower panel:} An image of the
  pulsar combining all visibilities in the observation, applying a
  correction for the dispersive time delay. The grayscale is linear
  from -42~Jy (white) to 1189~Jy (black). See \S\ref{s:mstransients}
  and \citet{ljb+11}.}
\label{f:pococrab}
\end{figure}

Figure~\ref{f:pococrab} demonstrates imaging of millisecond pulses
from the Crab pulsar (B0531+21) using a Python-based toolchain. We
observed the Crab pulsar at a time resolution of 1.2~ms and
frequencies between 720 to 800~MHz using the ATA \citep{ljb+11}. At
this time resolution, the dispersion introduced by the interstellar
medium (ISM) delays the pulse arrival time by a few tens of
milliseconds across the band. The upper panel of
Figure~\ref{f:pococrab} shows this effect in a spectrogram formed by
summing visibilities, effectively forming a synthesized beam toward
the Crab pulsar. The spectrogram shows that the dispersive delay
follows a quadratic shape that has a slope consistent with that
expected from the Crab pulsar. After identifying the time and
dispersion of the pulse, we can image it to see if it is consistent
with a point source in the location of the Crab pulsar. The
\uv\ input/output facilities of \mirpy\ are used to write a new
dedispersed visibility dataset. The lower panel of
Figure~\ref{f:pococrab} shows an image made from visibilities during
the pulse. The dispersive delay is large enough that it must be
corrected in order to detect the source.

\begin{figure}[htbp]
\plotone{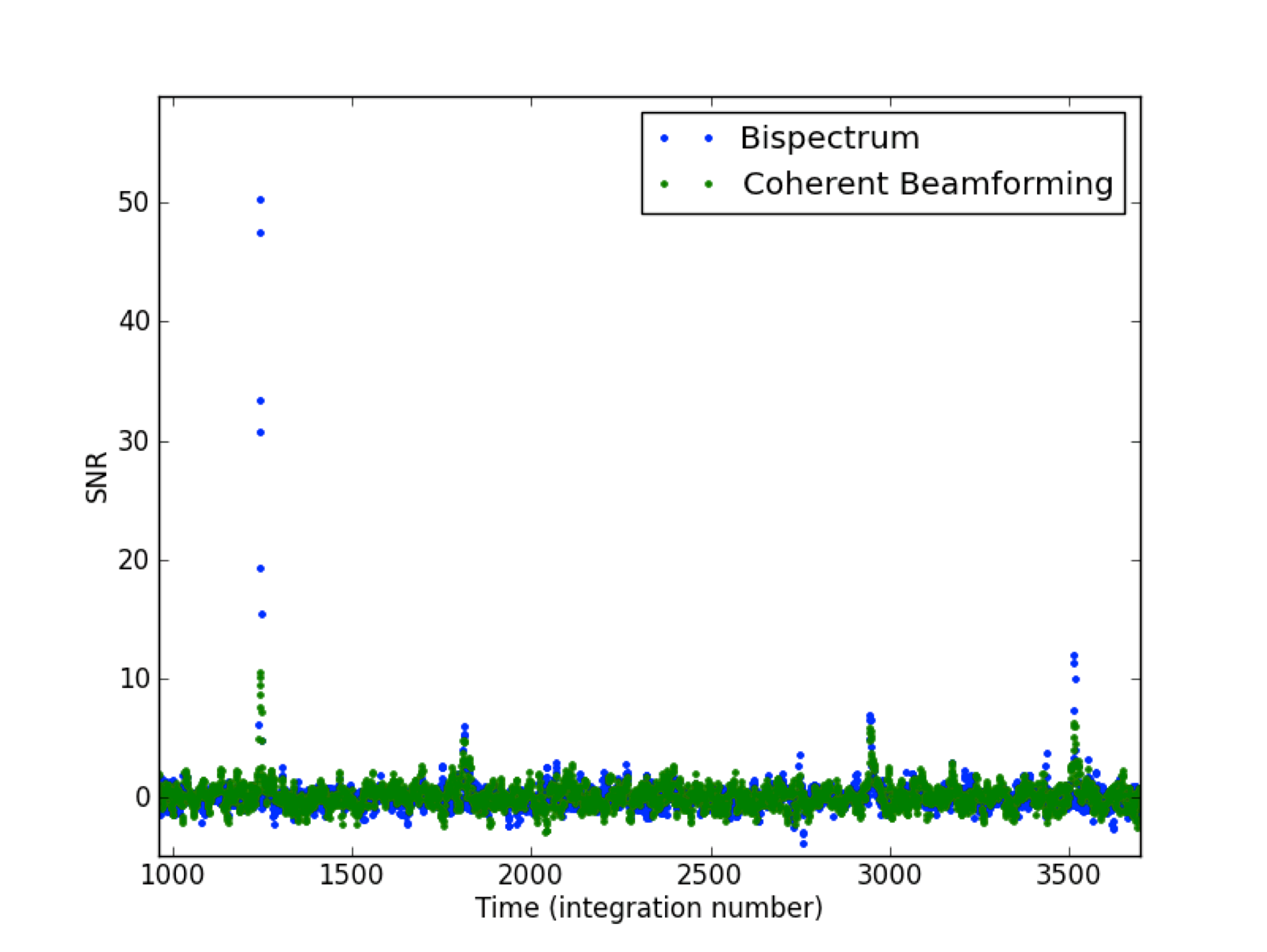}
\caption{Comparison of two methods for detecting millisecond
  astrophysical pulses, one using the bispectrum and one using
  traditional beamforming, both implemented in Python. Each method is
  applied to a visibility data stream containing five pulses from the
  bright pulsar B0329+54 and the signal-to-noise ratio (SNR) of each
  detection is plotted. The bispectrum-based method can be more
  effective (higher SNR detections for the same data) than coherent
  beamforming and is sensitive to pulses coming from any
  direction. See \S\ref{s:mstransients} and \citet{lb11arxiv}.}
\label{f:snrtpoco}
\end{figure}

Low-level access to visibility data in \mirpy\ also makes it possible
to experiment with new algorithms. We've used this capability to
develop a technique to detect millisecond transients based on an
interferometric closure quantity called the bispectrum
\citep{c87,lb11arxiv}. The bispectrum is formed by multiplying three
visibilities from baselines that form a closed loop. This product is
sensitive to transients anywhere in the field of view, which makes it
powerful for surveys.  However, most software packages only use the
bispectrum for calibration, so none have low-level access to functions
for prototyping algorithms. Figure~\ref{f:snrtpoco} shows millisecond
light curves toward pulsar B0329+54 made with \mirpy. With direct
access to dedispersed visibilities, we were able to compare
traditional beamforming with the bispectrum technique. In this case,
we can detect pulses for four of the five rotations of the pulsar
during this observation. Being able to compare the techniques on the
same pulses allowed us to show that the bispectrum responds more
strongly than beamforming, confirming an unusual theoretical property
of the bispectrum \citep{lb11arxiv}.

\subsection{Retroactive SEFD Calibration}

The ATA lacks online measurement of the system temperature
($T_\textrm{sys}$) or system equivalent flux density (SEFD),
information that is important for assessing data quality and thermal
noise limits. We have written a task, \textsf{calnoise}, to assess
per-antenna SEFDs after-the-fact using observations of a bandpass
calibrator and the assumption that the variance across the channels of
each spectral window is entirely thermal.

Each ATA antenna has two orthogonally linearly polarized feeds. The
signal from each of these is identified as originating from a given
antenna-polarization pairing (``antpol''). The SEFDs of the two feeds
on a single antenna may differ appreciably, and all of the SEFDs may
be time-variable. Rather than directly computing SEFDs,
\textsf{calnoise} determines calibration coefficients relating SEFDs
to the RMS value across the spectral window of each antpol's
uncalibrated autocorrelation amplitude, a value we refer to as the
RARA (raw autocorrelation RMS amplitude). While the value of this
coefficient is assumed to be static over the course of an observing
session, the derived SEFD value will vary proportionally with the
RARA. Although the RARA is a function of both system noise and a gain
factor, the time variation in the latter is small enough so that the
derived SEFDs will be sufficiently accurate.

\textsf{calnoise} takes as an input a dataset with a gains (and
optionally bandpass) table. The dataset is scanned through once,
without applying the gains table, to accumulate RARA values for each
timestamp and antpol. The dataset is then reread, with the gains being
applied, to compute per-baseline SEFDs: $$ \textrm{SEFD} =
\frac{1}{2}\left(\sigma_r + \sigma_i\right) \eta \sqrt{2 \Delta\nu
  \tau},$$ where $\sigma_r$ and $\sigma_i$ are the standard deviation
in the real and imaginary parts, respectively, of the calibrated
visibilities across the spectral window (thus in Jansky units), $\eta$
is a tabulated efficiency of the correlation (depending, e.g., on the
number of bits used in digitization), $\Delta\nu$ is the channel width
in the spectral window, and $\tau$ is the integration time of the
record. Per-antenna calibration coefficients are then calculated using
a least-squares fit of the baseline-based values assuming a simple
geometric dependence: $$\textrm{SEFD}_{i,j,t} = \sqrt{c_i
  \textrm{RARA}_{i,t} \cdot c_j \textrm{RARA}_{j,t}},$$ where the
$c_i$ are the desired calibration parameters. Outlier values (possibly
due to RFI or hardware failures) are identified by user-specified
clipping limits and iteratively removed from the fits. MIRIAD's
on-the-fly calibration system does not support SEFD information, so
the $c_i$ are then recorded in a textual table on disk; a companion
task reads the table along with an input dataset and creates a new
dataset with the SEFD information inserted. (For technical reasons, a
constant $T_\textrm{sys}$ is assumed and varying values of the MIRIAD
variable \textsf{jyperk} are inserted into the dataset, where
$\textrm{SEFD} = T_\textrm{sys} \cdot \textsf{jyperk}$ and thus
\textsf{jyperk} is related to the effective area of each receiving
element.)

\textsf{calnoise} is written in Python and uses \mirpy\ to read the
visibility data and for miscellaneous astronomical routines. It uses
NumPy for its numerics and other Python libraries for optional
plotting of diagnostic information, least-squares solving, storage of
numerical metadata, and integration into the ATA pipeline described
below (\S\ref{s:atapipeline}). The main implementation comprises \apx
600 statement lines of code (i.e., excluding whitespace and
comments). The chief benefits of using Python and \mirpy\ were quick
creation of the task skeleton (thanks mainly to high-level data
structures), rapid turnaround during refinement (thanks mainly to the
lack of a compilation step), and easy investigation of algorithmic
tweaks (thanks mainly to easy access to other libraries, e.g., only
one line of code needed to interactively plot variables).

\subsection{Interactive \uv\ Data Visualization}
\label{s:uvvis}

\begin{figure}[htbp]
\plotone{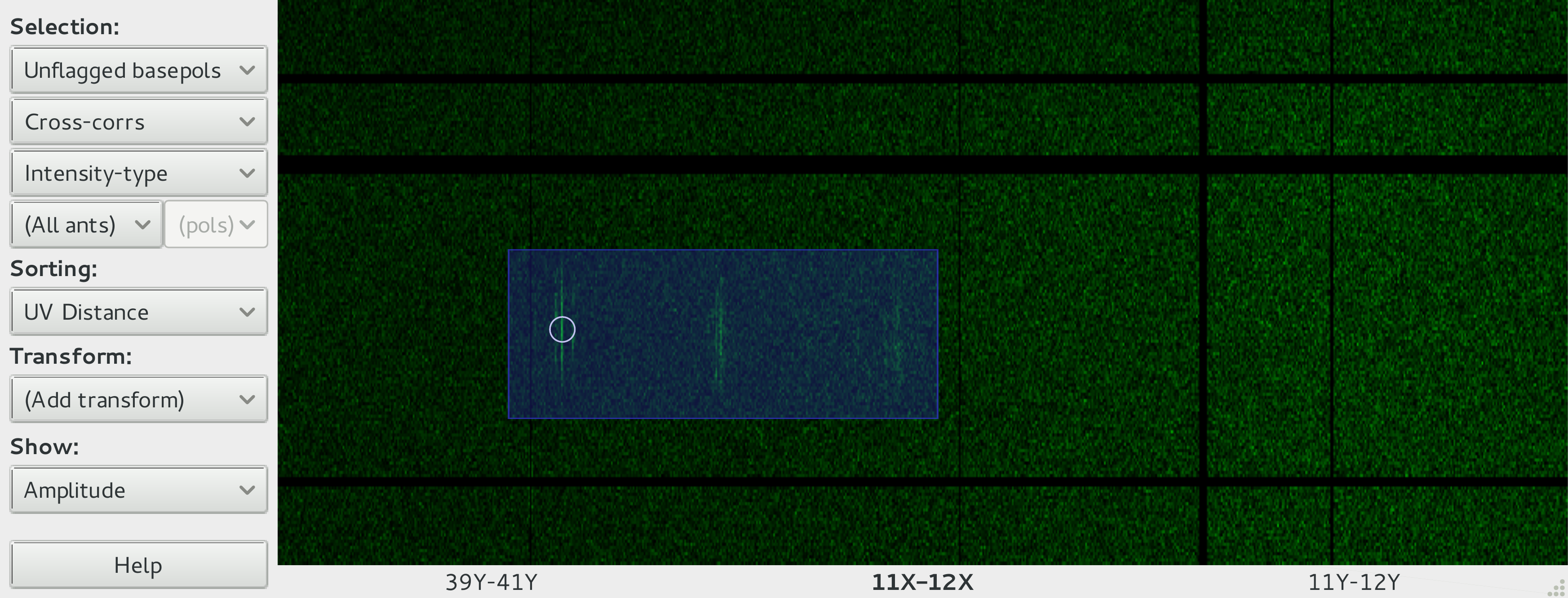}
\caption{Sample screenshot of the \uv\ data visualizer described in
  \S\ref{s:uvvis}. The main graphical display is a dynamic spectrum of
  amplitude on one baseline as a function of frequency (horizontal
  axis) and time (vertical axis). The left-hand panel provides
  controls for filtering and transforming the data. Python bindings to
  well-established graphical toolkits allow the data to be displayed
  in an attractive and responsive user interface.}
\label{f:uvshot}
\end{figure}

Data visualization is an important part of the development of any
processing pipeline. This is especially true for radio
interferometers, in which the data undergo complex transformations,
are high-dimensioned, and closed-loop instrumental modeling is often a
key aspect of the reduction. We have used \mirpy\ to implement a
powerful \uv\ data visualizer, first described in \citet{w10}. As
shown in Fig~\ref{f:uvshot}, the main display is a dynamic spectrum of
the visibilities on a baseline as a function of frequency and
time. The user can quickly switch between different displays (real,
imaginary, amplitude, phase), apply various processing steps
on-the-fly (e.g., average, rephase), and navigate through the dataset.
There is also substantial support for visualizing and creating data
flags when necessary, although it is widely recognized that manual
flagging of data is rapidly becoming an impossible task as data rates
increase \citep{kbw10}.

The use of Python is a key aspect to the \uv\ visualizer because it
allows easy combination of the existing MIRIAD libraries with very
different software, in this case the modern graphical toolkits
Cairo\fnu{http://cairographics.org/} and
GTK+\fnu{http://gtk.org/}. Cairo provides routines for fast rendering
of the gridded visibility data, while GTK+ provides a higher-level
widget library that allows new user interactions to be implemented
quickly. Achieving this functionality without the use of these (or
similar) libraries would be a massive undertaking: Cairo and GTK+
comprise 240000 and 560000 lines, respectively, of well-tested,
efficient C code. The main visualizer codebase, on the other hand,
comprises only 2900 statement lines of Python.

\subsection{An ATA Reduction Pipeline}
\label{s:atapipeline}

Modern and next-generation radio observatories are expected to produce
data at rates significantly higher than those of older facilities
\citeeg{ch10b}. Under these conditions, rapid automated processing
of data is changing from a luxury to a necessity \citep{kbw10}. Python
is often identified as a good system in which to construct pipelines
to perform such processing, thanks to its amenability to implementing
high-level application logic, revisability, and ability to glue
together existing tools and libraries \citep{s99,mgs07,pgh11}. We have
used \mirpy\ to implement a commensal observing system \citep{w12} for
the ATA Galactic Lightcurve and Transient Experiment (AGILITE;
Williams et~al., 2012, in prep.) and just such a pipeline for
processing the data. In the latter case, the \mirexec\ module is used
extensively to invoke existing MIRIAD tasks and \mirtask\ is used for
low-level examination and manipulation of the data, often to
accomplish tasks specific to ATA data reduction.

Of particular note is that the pipeline internally manages data flow
through a directed acyclic graph model analogous to that of the
traditional Unix tool \textsf{make} \citep{themake}. While traditional
shell scripting languages do not support the data structures necessary
to conveniently implement such a design, doing so is easy in Python.
The \textsf{drPACS} package \citep{thedrpacs} takes a similar approach
but actually uses \textsf{make} for its underlying management. We find
this approach to be unsatisfactory for two reasons. Firstly,
\textsf{make} can only track dependencies between files on disk, and
can only invoke tools via the Unix shell. This model maps very
inefficiently onto certain aspects of typical data reduction
workflows. Secondly, and more fundamentally, \textsf{make} checks
whether steps need to be rerun by comparing file modification times
rather than actual contents. This suffers from a variety of minor
issues and the major issue that if a pipeline product is regenerated,
all products downstream of it must also be regenerated, even if the
rebuilt product did not actually change. In a data reduction pipeline
the costs of this inefficiency can be severe. The tool described here
avoids this problem by detecting modifications with cryptographic
hashes as alluded to in \S\ref{s:design}.

\section{\mirpy\ as a Software Project}
\label{s:project}

In this section we describe \mirpy\ as a software project. We also
hope that this section may provide future authors a useful reference
of issues that, in our opinion, are important to discuss when
documenting any other software project.

\subsection{Intellectual Property Issues}

\mirpy\ is an open source project, almost entirely licensed under the
GNU General Public License, version 3 or later. Small portions of it
relating to the compilation process are licensed under different
permissive licenses. As such, \mirpy\ is available for inspection and
use by anyone for any purpose. The source code copyright is owned by
the authors. \mirpy\ is not known to be subject to any software
patents.

\subsection{Availability}

The source code is available for free download from the project
website\footnote{\url{http://purl.org/net/pkgwpub/miriad-python}. This
  link is a ``permanent URL'' providing a durable, reroutable link
  suitable for inclusion in the academic literature. The authors
  encourage the use of this URL and not its current,
  possibly-ephemeral destination when linking to the \mirpy\ website
  and related pages. See \url{http://purl.org/} for more information.}
either in the form of file archives or via the
\git\fnu{http://git-scm.com/} version control system. The
\git\ repository is also available on the website
GitHub\footnote{Currently
  \url{https://github.com/pkgw/miriad-python/}, but the canonical link
  may be found via the project website.}. Thanks to the distributed
nature of \git, clones of either repository stand alone and contain
the complete, cryptographically-verified revision history of the
entire project\footnote{The SHA1 checksum of the version of
  \mirpy\ described in this paper is \cursha. The secure architecture
  of \git\ ensures that a valid commit of this checksum is guaranteed
  to correspond to the exact source code described in this
  work.}. Most installations of \mirpy\ track the \git\ repository and
so official releases are rare. When made, they are documented and
linked to on the project website. The most recent release is version
\currelvers, made on \curreldate\footnote{The corresponding
  \git\ commit is \currelsha.}.

Installation instructions are beyond the scope of this paper and are
provided on the website. We note, however, that \mirpy\ must be
compiled against MIRIAD libraries from the CARMA codebase built with
the \textsf{autoconf}-based build system.

\subsection{Development Model}

Development of \mirpy\ occurs in the \git\ repository. The authors aim
to conduct development in an open, welcoming fashion. Contributions
from the community are encouraged via (e.g.)  email or ``pull
requests'' on the GitHub site. Based on the experiences of other
community-based software projects, copyright assignment is not
required for external contributions \citep[see, for instance, the
  discussion in][]{om03}.

The scale of the \mirpy\ project is such that other pieces of project
coordination infrastructure such as mailing lists and a bug tracker
are not currently deemed necessary. These will arise as the
\mirpy\ community grows, and will be linked to on the project website.

\subsection{Documentation and Examples}
\label{s:docs}

A moderately-complete manual to \mirpy\ is available on the website,
providing both an API reference and somewhat higher-level guidance as
to the intended usage of the package. The primary documentation format
is HTML served over the web, but the documentation is generated using
Sphinx\fnu{http://sphinx.pocoo.org}, the official documentation system
of the Python language, and in principle the documentation can be
generated in several other output formats, including printable PDF
files.

Besides the brief examples in \S\ref{s:examples}, the \mirpy\ source
distribution contains a small set of longer samples implementing
features as both standalone scripts and importable modules. Exercised
features include reading and writing of UV data, executing MIRIAD
tasks with \mirtask, manipulating dataset header items, reading gains
tables, and integrating into the MIRIAD documentation system.

\subsection{Quality Assurance}
\label{s:qa}

\mirpy\ is in line with the vast majority of scientific software in
that, unfortunately, it has virtually no quality-assurance systems in
place beyond the fact that it is exercised in day-to-day use. In
particular, it has no automated test suite. Code changes in
\textsf{diff} format are reviewed before being committed to the
repository, hopefully preventing unintended changes from being
integrated into the source tree. An automatic test framework is
planned, using short unit tests of the \mirpy\ APIs as well as more
involved tests using sample MIRIAD datasets either generated
on-the-fly (with \textsf{uvgen}) or optionally downloaded from the
project website.

Given time constraints and the potential payoffs it is certainly not
hard to understand why scientific software so often lacks systematic
testing infrastructure. As a point of reference, testing can approach
half the cost of a piece of software in the corporate context
\citep{mvm10}. We emphasize, however, that software defects are
extraordinarily common --- about one for every {\it twenty} statement
lines of code for undisciplined coding \citep{sbb+02} --- and that
even rudimentary testing can capture many of these defects
\citeeg{b96}. \citet{sbb+02} present an excellent summary of the
prevalence and impact of defects in typical software systems.

\section{Performance Considerations}
\label{s:perf}

Python, being a dynamic interpreted language, is not optimal for
achieving raw numerical throughput. Indeed, the {\it raison d'\^etre}
of a tool such as \mirpy\ is that in many cases numerical throughput
can profitably be traded for other desirables such as code
readability, programmer time savings, and ease of integration with
existing codebases. In our experience, data-processing algorithms
written in Python are almost always ``fast enough,'' even for fairly
complex algorithms.

\subsection{Serial Tests}

To give a very rough quantification of the overhead of using Python
compared to FORTRAN-77, we wrote two versions of a simple task that
iterates through a MIRIAD \uv\ dataset and computes the RMS value of
the unflagged visibilities in each record. One version was written in
Python and one in FORTRAN-77, the latter being compiled with GNU
\textsf{gfortran} version 4.4.1. For a 0.3 GB visibility dataset
containing 2,812,072 \uv\ records of 16 channels each, the Python
version ran about 21 times slower than the FORTRAN-77 version (63~s vs
3~s) on our test system, a machine running a quad-core AMD
Opteron~1385 CPU at 2.7~GHz. (Tests were run repeatedly with no effort
to flush filesystem caches, so performance was CPU-limited.) For a 1.6
GB dataset containing 406,260 \uv\ records of 1,024 channels each, the
Python version ran about 4 times slower (26~s vs 7~s). These results
suggest that there is indeed a nonnegligible overhead to the Python
interpreter, although its significance can be highly dependent on the
structure of the input data. The performance of the Python version in
the 1,024-channel case is relatively good because relatively more work
is done inside a few vectorized NumPy function calls; manually
iterating over each channel in Python increases the runtime by a
factor of \apx 90. For cases where most of the work is performed
inside a few vectorized NumPy functions, the overhead to using Python
can be fairly small.

In most software, the majority of the execution time is spent in only
a small portion of the code. For those cases where serial throughput
is a limitation, large improvements can often be achieved for low
effort by identifying the most-executed codepaths and porting their
particular implementations to a compiled language. This strategy is
adopted by several of the projects discussed in \S\ref{s:compare}. In
some instances, it may be possible and profitable to generate
speed-critical code on-the-fly using a compiler library such as the
Low-Level Virtual Machine \citep{thellvm}, avoiding the use of
customized precompiled modules altogether. In other cases, it may make
sense to prototype an algorithm in Python and then develop a
production implementation in a compiled language. Of course, there
will always exist problems for which Python is an inappropriate tool,
but in our experience the combination of Python and a few basic
numerical libraries is sufficient to address a wide range of
challenges.

\subsection{Parallelization}

It is clear that parallelized processing will play a fundamental role
in the reduction of future interferometric datasets. There is a
special opportunity for next-generation interferometric processing
packages to establish themselves in this space because the existing
packages were generally built purely for serial processing.

There are several classes of parallelization that may be
considered. For some applications, multiple independent reduction
processes may be executed simultaneously on a single machine or on a
cluster with no intra-process communication or synchronization. In
this ``embarrassingly parallel'' case, a language such as Python may
ease the development of both the reduction processes and the system to
manage the overall processing job. Indeed, there are several
preexisting Python frameworks for constructing parallel applications,
e.g. Parallel Python\fnu{http://www.parallelpython.com/}.

Other parallel applications can be implemented via multiple
independent processes that communicate and synchronize. Such
applications are often implemented with the standard Message Passing
Interface \citep[MPI;][]{thempi}, which is accessible in Python via
several different toolkits \citep[e.g.,
  \textsf{mpi4py};][]{thempi4py}. Somewhat surprisingly, MPI-based
parallel Python programs with core numerical routines written in a
compiled language can perform as well as parallel programs written
purely in a compiled language \citep{clm05}. The comparative ease of
writing the driving logic of such a program in Python versus compiled
languages makes this an intriguing model for parallel algorithmic
development.

For applications that require more intensive data-sharing, one may
wish to parallelize MIRIAD operations within a single process by using
multithreading. Unfortunately, Python's ``global interpreter lock''
prevents multiple threads from actually executing concurrently. (In
many cases, na\"ively multithreading a Python program makes it much
slower!) In the particular case of \mirpy, extreme care would also be
needed to synchronize access to the MIRIAD subroutine libraries, which
are not threadsafe and maintain substantial shared state. \citet{c08}
describes some of the challenges and successes encountered in tackling
this kind of problem in Obit. The recommended system for
threaded-style computation in Python is the \textsf{multiprocessing}
module, which provides an infrastructure for launching concurrent
Python subprocesses and communicating objects between parent and
child. In the context of this work, \textsf{multiprocessing} is a more
limited, Python-specific variant of the standard MPI approach. For
certain I/O-limited applications, multithreaded Python code can obtain
a performance gain above single-threaded code, but the difficulties of
serializing access to the MIRIAD libraries remain.

Finally, certain classes of problems are well-suited to the
highly-parallel capabilities of GPUs (graphics processing
units). Current uses of GPUs in interferometric applications include
pulse dedispersion \citeeg{mks+11,thedspsr}, RFI mitigation
\citep{aawdv+12}, cross-correlation \citep{cpg11arxiv}, and
visualization \citep{hfb11}. Python code can take advantage of GPUs
using frameworks such as PyCUDA \citep{thepycuda}. The nature of these
examples suggest that the algorithms most suited to GPU processing
would not likely need to build on the MIRIAD/\mirpy\ infrastructure,
except perhaps to use MIRIAD data formats for I/O.

Regardless of the kind of parallelization in question, many
interferometric algorithms are I/O-intensive. The options for
improving the I/O bandwidth of MIRIAD/\mirpy\ are somewhat limited due
to the lack of built-in support for parallel I/O \citeeg{theromio} in
the MIRIAD I/O subroutines. With investment in the appropriate
hardware, speedups can be achieved using transparently parallel
filesystems \citep[e.g., Lustre\fnu{http://lustre.org/} or
  PVFS;][]{thepvfs}. Significant gains in effective I/O bandwidth may
alternatively be achieved with commodity hardware by spreading data
between many independent processing nodes if the application permits
partitioning of the data (e.g., by spectral channel or pointing
direction).

\section{Summary}
\label{s:summary}

The new generation of interferometric hardware demands a new
generation of interferometric software implementing a new generation
of algorithms. Fortunately, there's a wide variety of projects aiming
to provide the infrastructure needed to develop these tools as well
as going ahead and implementing them.

\mirpy\ is one of these projects. It is ``broad'' rather than
``tall'': it provides a wide range of APIs for accessing MIRIAD tasks
and data, but does not provide its own algorithms built on top of
those APIs. While certain applications will be best matched to the
facilities provided by the higher-level packages Obit, AIPY, or
MeqTrees, \mirpy\ is a good foundation on which to build those that
are not, especially if one wishes to take advantage of the simple and
efficient MIRIAD \uv\ data format. We actively encourage contributions
to \mirpy\ development and are interested in supporting new
\mirpy\ applications.

It seems likely that SKA-scale interferometric software will be only
tangentially, if at all, related to any of the packages that exist
today, and it is unclear if humans will ``reduce'' SKA data in any
meaningful way given the rates involved. If the SKA is to meet both
its goals and its schedule, however, the techniques needed to conquer
exascale data rates must be discovered and prototyped, piece by piece,
using today's software packages. Experience suggests that the
innovations needed to solve these kinds of massive challenges do not
come out of one organization; instead, wide-ranging and vibrant
experimentation by a broad community of contributors will be
essential.

\acknowledgments

The authors thank W.~D.~Cotton for helpful discussions. \mirpy\ is
built upon the open-source Python language as well as an extensive
foundation of other open-source and Free Software packages. It could
not exist without the work of the innumerable people whose
contributions have helped build this infrastructure. Research with the
ATA is supported by the Paul G. Allen Family Foundation, the National
Science Foundation, the US Naval Observatory, and other public and
private donors. This research has made use of NASA's Astrophysics Data
System.

\bibliographystyle{yahapj}
\bibliography{pkgw}{}

\end{document}

%% file: figaliases.tex
\newcommand\figblockdiagram{f1}
\newcommand\figpococraba{f2a}
\newcommand\figpococrabb{f2b}
\newcommand\figsnrtpoco{f3}
\newcommand\figuvshot{f4}